\documentstyle[aps,prl,epsfig]{revtex}

\begin{document}

\title{Non-Abelian Thermal Large Gauge Transformations in 2+1 Dimensions}

\author{F. T. Brandt$^a$, Ashok Das$^b$, Gerald V. Dunne$^c$,
J. Frenkel$^a$ and  J. C. Taylor$^d$}
\address{$^a$Instituto de F\'{\i}sica,
Universidade de S\~ao Paulo,
S\~ao Paulo, SP 05315-970, BRAZIL}
\address{$^b$Department of Physics and Astronomy,
University of Rochester,
Rochester, NY 14627-0171, USA}
\address{$^c$Department of Physics,
University of Connecticut,
Storrs, CT 06269-3046, USA}
\address{$^d$ Department of Applied Mathematics, 
and Theoretical Physics,
University of Cambridge,
Cambridge, UK}

\maketitle
\vskip .5cm

\begin{abstract}
We discuss several different constructions of non-Abelian {\em large} gauge
transformations at finite temperature. Pisarski's ansatz with even
winding number is related to Hopf mappings, and we present a simple new
ansatz that has any integer winding number at finite temperature.
\end{abstract}

\section{Introduction}

The question of {\em large} gauge invariance at finite temperature
has led to a lot of activity in recent years. The basic problem is that,
while finite temperature perturbation theory is invariant order by
order under {\em small} gauge transformations, it is not invariant, order by
order, under {\em large} gauge transformations. The nontrivial
periodicity conditions satisfied by finite temperature {\em large} gauge
transformations mix all orders of perturbation theory. This phenomenon
leads to complications in computing finite temperature effective actions
when there is an anomalous symmetry such as parity that can produce
anomalous terms in the effective action. This problem has mostly been
studied in the context of finite temperature Chern-Simons theories, both
in $0+1$ and $2+1$ dimensions \cite{dunne,deser1,schaposnik,lhl}. In
Chern-Simons theories this is a serious problem, because a non-Abelian
Chern-Simons term shifts by a constant under a {\em large} gauge
transformation. At zero temperature a consistent quantum theory can still
be defined if the coefficient of the Chern-Simons term takes discrete
values \cite{deser2}, and, remarkably, these discrete coefficients are
precisely what is found for Chern-Simons terms that are induced by
radiative quantum loops \cite{redlich,pr}. At finite temperature the
induced coefficient is temperature dependent and so does not take discrete
values, suggesting the possibility of a violation of {\em large} gauge
invariance \cite{pisarski}. The resolution of this problem has now been
understood when the {\em large} gauge transformation is essentially Abelian in
the sense that it is the nontrivial winding of an Abelian field around
the Euclidean time ${\bf S}^1$ that produces the shift in the Chern-Simons
term \cite{dunne,deser1,schaposnik,lhl}. Very briefly, at finite
temperature, the effective action consists of an infinite number of
parity-violating terms, of which the Chern-Simons term is only the first,
in such a way that the complete effective action is invariant under {\em
large} gauge transformations, although a truncation of the effective
action at any order leads to violation of this symmetry. However, this
mechanism has only been explicitly verified for a very restricted type of
Abelian {\em large} gauge transformations. The situation for a genuinely
non-Abelian {\em large} gauge transformation, whose non-vanishing
winding  number
comes from its mapping from ${\bf S}^2\times {\bf S}^1$ into the
non-Abelian  gauge
group, has not yet been fully understood. However, explicit computations of
the finite temperature multi-leg amplitudes in non-Abelian theories
indicate that the structure of the non-Abelian parity-odd parts is much
richer at finite temperature than at zero temperature or in finite
temperature Abelian theories \cite{bdf}.

Clearly, an important part of this entire discussion is to understand the
properties of the finite temperature {\em large} gauge transformations
themselves. Here one strikes immediately a conceptual difference between
zero and nonzero temperature. For simplicity, we consider the gauge
group to be $SU(2)$. The group manifold $SU(2)$ can be identified with
${\bf S}^3$. At zero temperature, the winding can be thought of as resulting
from a map of space-time to the group space,
${\bf S}^{3}\rightarrow {\bf S}^{3}$. These windings take integer values
since the homotopy group is $\pi_3(S^3)={\bf Z}$. On the other hand, at
finite temperature the space-time manifold becomes
${\bf S}^{2}\times {\bf S}^{1}$, in the imaginary time formalism of
thermal field theories. Thus, at finite temperature, homotopy groups are
not sufficient to characterize these maps, since both $\pi_1(S^3)$ and
$\pi_2(S^3)$ are trivial. This is because there can only
be trivial maps between ${\bf S}^{m}\rightarrow {\bf S}^{n}$, when $m <
n$. Therefore, it might appear naively that there can be no {\em
large} gauge transformations at finite temperature. This naive
expectation is not correct; indeed an explicit ansatz for finite
temperature {\em large} gauge transformations has been presented by Pisarski
\cite{pisarski}. However, this ansatz is restricted to {\it even} winding
numbers only. Furthermore, it does not have a smooth zero temperature
limit to a zero temperature {\em large} gauge transformation. 

In this paper, we
explore the geometric properties of these finite temperature {\em large} gauge
transformations, and discuss several other explicit ansatzes. Some of
these also have only even winding number, but we also present a very
simple ansatz for a finite temperature {\em large} gauge transformation which
has any integer winding number and which has a smooth zero T limit. In Section
II, we briefly recall the zero temperature {\em large} gauge
transformations. In
Section III, we discuss Pisarski's ansatz that leads to genuinely
non-Abelian {\em large} gauge transformations with even winding number. This
construction is shown, in Section IV, to be related to Hopf maps. In
Section V,  we
show how a coset construction, at finite temperature, gives only trivial
winding. In Section VI, we show how one can enlarge the group space to
have a nontrivial map from ${\bf S}^{2}\times {\bf S}^{1}$. However, such
a construction also leads to only even winding numbers. In Section VII,
we present an alternate construction, borrowing from ideas in $0+1$
dimensions, that leads to an arbitrary winding number. Section VIII
contains some brief conclusions and we collect some useful formulae
about winding numbers in an appendix.

\section{Zero temperature}

At zero temperature, a simple ansatz for an $SU(2)$ {\em large} gauge
transformation is \cite{roman1}
\begin{eqnarray}
g(\vec{x})=\exp\left(\frac{m\pi
i\,\vec{x}\cdot\vec{\sigma}}{\sqrt{\vec{x}^2+\lambda^2}}\right)
\label{zt}
\end{eqnarray}
where $\vec{x}\in {\bf R}^3$, $\vec{\sigma}$ are the Pauli matrices, $m$ is
an integer, and
$\lambda$ is an arbitrary scale parameter. For later comparison, it is
convenient to express $g$ as
\begin{eqnarray}
g(\vec{x})=\cos\left(\frac{m\pi
r}{\sqrt{r^2+\lambda^2}}\right) {\bf 1} +i\,(\hat{x}\cdot\vec{\sigma})\,
\sin\left(\frac{m\pi r}{\sqrt{r^2+\lambda^2}}\right)
\label{zttrig}
\end{eqnarray}
where $r^{2} = \vec{x}^{2}$, and  $\hat{x}=\frac{\vec{x}}{r}$ is the
three-dimensional unit position vector. Note that
\begin{eqnarray}
g(r=0)= {\bf 1} \quad ,\quad g(r=\infty)=\cos(m \pi)\, {\bf 1}
\label{ztlimits}
\end{eqnarray}
The winding number of an $SU(2)$ group element is 
\begin{eqnarray}
W[g]  =  {1\over 24\pi^{2}}  \int d^{3}x\,\epsilon_{\mu\nu\lambda} {\rm
Tr} \left( g^{-1}\partial_{\mu}g
g^{-1}\partial_{\nu}g g^{-1}\partial_{\lambda}g \right)
\label{winding}
\end{eqnarray}
It is a straightforward exercise to verify that, for the zero
temperature ansatz (\ref{zt}), the winding number of $g$ is equal to the
integer $m$:
\begin{eqnarray}
W[g]=m
\label{ztm}
\end{eqnarray}
Geometrically, $W[g]$ is the number of times the map $g:{\bf S}^3\to
{\bf S}^3$ winds
around the target manifold $SU(2)\sim {\bf S}^3$ as the coordinates cover the
base manifold ${\bf R}^3$ (which can be compactified to ${\bf S}^3$ if
$g$  has a
limit that is independent of the angles at $r=\infty$). In other words,
the integrand in (\ref{winding}) is just ($m$ times) the Jacobian
involved in changing variables from Cartesian coordinates to the angular
coordinates used to parameterize ${\bf S}^3$ (see, e.g.,
\cite{rajaraman}).

\section{Finite Temperature : Pisarski's Ansatz}

At finite temperature, in the imaginary time formulation, a gauge
transformation $g(\bar{x},t)$ must be strictly periodic \cite{comment} in
the Euclidean time $t$, with period $\beta=\frac{1}{T}$ being the inverse
temperature \cite{kapusta,lebellac,dasbook}:
\begin{eqnarray}
g(t=0)=g(t=\beta)
\label{periodic}
\end{eqnarray}
Since the coordinate manifold is now ${\bf S}^2\times {\bf S}^1$ rather than
${\bf S}^3$, it is natural to separate the $t$ dependence. Pisarski
\cite{pisarski} proposed the following elegant ansatz for a finite
temperature $SU(2)$  {\em large} gauge transformation:
\begin{eqnarray}
g(\bar{x},t) &=& \exp\left({2m\pi i t\over \beta}\,
\hat{n}(\bar{x})\cdot
\vec{\sigma}\right)\nonumber\\
&=&\cos\left(\frac{2m\pi
t}{\beta}\right) {\bf 1} +i\,(\hat{n}\cdot\vec{\sigma})\,
\sin\left(\frac{2 m\pi t}{\beta}\right)
\label{pa}
\end{eqnarray}
where $m$ is an integer, and $\hat{n}(\bar{x})$ is a three-component unit
vector depending only on the two-component spatial coordinate vector
$\bar{x}\in {\bf R}^2$. If we require that $\hat{n}(\bar{x})$ have an
angle-independent limit at spatial infinity ($|\bar{x}|=\infty$), then
${\bf R}^2$ is compactified to ${\bf S}^2$, and $\hat{n}$ defines a map
$\hat{n}:{\bf S}^2\to {\bf S}^2$. A straightforward calculation using
the  ansatz (\ref{pa}) for
$g(\bar{x},t)$ shows that the winding number (\ref{winding}) of $g$ is
related to the index of the map $\hat{n}$ in the following simple manner:
\begin{eqnarray}
W[g]=2m\, w[\hat{n}]
\label{even}
\end{eqnarray}
where
\begin{eqnarray}
w[\hat{n}]=\frac{1}{8\pi}\int d^2 x\, \epsilon^{abc}\epsilon^{ij}
\hat{n}^a \partial_i \hat{n}^b \partial_j \hat{n}^c
\label{2dwinding}
\end{eqnarray}
This winding number of $\hat{n}$ is itself an integer, which implies (as
noted in \cite{pisarski}) that the winding number of $g$ with the ansatz
(\ref{pa}) is necessarily an {\it even} integer.

Explicit examples of $\hat{n}(\bar{x})$ with nontrivial integer windings
as maps from ${\bf S}^2$ to ${\bf S}^2$ are provided by the $CP^1$ instantons
\cite{rajaraman}. For example, the map (here $z={x^1+ix^2\over \lambda}$)
\begin{eqnarray}
\hat{n}=\frac{1}{1+|z|^2}\left(\matrix{2Re(z)\cr 2 Im(z)\cr
1-|z|^2}\right)
\label{cp1}
\end{eqnarray}
has $w[\hat{n}]=1$. Geometrically, this can simply be viewed as the
stereographic projection connecting the unit sphere ${\bf S}^2$ to the
plane.  If the ${\bf S}^2$ is embedded into ${\bf R}^3$, then we can
write this as $\hat{n}=\hat{x}$.

Clearly, the ansatz (\ref{pa}) is manifestly periodic in Euclidean time,
with period $\beta$. Further, it gives a nonzero winding number
(\ref{even}). Thus, it is indeed a genuine non-Abelian finite temperature
{\em large} gauge transformation. However, this construction also
raises  several
questions. First, why are only even winding numbers allowed? Second, what
is the geometrical interpretation of this construction in terms of
mappings from ${\bf S}^2\times {\bf S}^1$ to $SU(2)$? Third, why is it
that  this ansatz does not survive the zero temperature
($\beta\to\infty$) limit and reduce to the zero temperature {\em
large} gauge transformation in (\ref{zt})?

A partial answer to the first question can be given as follows. Notice
that if we consider the two-dimensional
instanton in (\ref{cp1}) for which we can identify $\hat{n}$ with the
three-dimensional unit vector $\hat{x}$, then the finite T ansatz
(\ref{pa}) is very similar to the zero temperature ansatz (\ref{zttrig}).
The only difference is that in the zero T case, as $r$ goes from $0$ to
$\infty$, the trigonometric factors go through $m$ half cycles, while in
the finite T case, as $t$ goes from $0$ to $\beta$, the trigonometric
factors go through $m$ full cycles. This difference is forced by the
strict periodicity condition at finite temperature, and means that the
finite T ansatz is wrapping twice as often around $SU(2)$. A deeper
geometrical interpretation of this phenomenon is presented in the next
section. Let us also point out here that Eq. (\ref{pa}), considered as
a map from space-time to the group manifold, has no inverse, because
$g(\bar{x},0) = {\bf 1}$ (for all positions). This is an important
difference from the zero temperature case. The same behavior holds for
the other examples of thermal {\em large} gauge transformations in
this paper.

\section{Relation to Hopf maps}

With every $SU(2)$ group element, $g:{\bf R}^3\to SU(2)$, it is possible
to associate a three-component unit vector $\hat{N}\in {\bf S}^2$ in such a way
that the winding number of $g$ (viewed as a map from ${\bf S}^3\to
{\bf S}^3$) is
equal to the Hopf index of $\hat{N}$ (viewed as a map from $S^3\to S^2$).
[It is important not to confuse $\hat{N}(\vec{x})$ with 
$\hat{n}(\bar{x})$ of the previous section. $\hat{N}$ maps from a 
three-dimensional base manifold into ${\bf S}^2$, while $\hat{n}$ mapped from a
two-dimensional base manifold into ${\bf S}^2$.] This Hopf construction has
been intensely studied recently in the physics literature in a wide
variety of contexts ranging from knot solitons \cite{niemi,paul,tsutsui},
to  magnetic helicity \cite{roman}, to zero modes of Abelian Dirac
operators \cite{adam}, to Abelian projections \cite{jahn,vanbaal}. The
specific relation between $g$ and $\hat{N}$ is
\begin{eqnarray}
\hat{N}^a=\frac{1}{2}tr\left(\sigma_3\, g^{-1}\sigma_a g\right)
\label{nhat}
\end{eqnarray}
It is simple to verify that $\hat{N}^2=1$, so that $\hat{N}\in {\bf S}^2$.
Algebraically, $g$ is the local unitary transformation that diagonalizes
$\hat{N}\cdot \vec{\sigma}$ to its asymptotic value (at spatial
infinity) of $\sigma_3$. The Hopf index of $\hat{N}$ is 
\begin{eqnarray}
H[\hat{N}]=\frac{1}{8\pi^2} \int d^3 x\, \epsilon^{ijk}a_i f_{jk}
\label{hopf}
\end{eqnarray}
where $a_j$ is an associated Abelian gauge field (connection)
\begin{eqnarray}
a_j=-\frac{i}{2}tr\left(\sigma_3 g^{-1}\partial_j g\right)
\label{abelian}
\end{eqnarray}
and $f_{jk}=\partial_j a_k-\partial_k a_j$ is the corresponding Abelian
field strength (curvature). Then it is straightforward to show that the
Hopf index (\ref{hopf}) of $\hat{N}$ is equal to the winding number
(\ref{winding}) of $g$:
\begin{eqnarray}
H[\hat{N}]=W[g]
\label{hw}
\end{eqnarray}
The geometrical interpretation of the Hopf index is as follows
\cite{paul,jahn}. The pre-image of any fixed point $\hat{N}$ on ${\bf
S}^2$ describes a closed loop in ${\bf R}^3$. The Hopf index of $\hat{N}$
is the linking number of the pre-image loops for any two points on ${\bf
S}^2$.

Now consider the zero and finite temperature {\em large} gauge transformations
(\ref{zt}) and (\ref{pa}), respectively, in this Hopf map language. For
the zero T group element in (\ref{zt}), the relation (\ref{nhat}) leads to
\begin{eqnarray}
\hat{N}^a=\cos\left(\frac{2m\pi
r}{\sqrt{r^2+\lambda^2}}\right)\delta^{a3} +
\left[1-\cos\left(\frac{2m\pi r}{\sqrt{r^2+\lambda^2}}\right)\right]
\hat{x}^a \hat{x}^3 -
\sin\left(\frac{2m\pi r}{\sqrt{r^2+\lambda^2}}\right)
\epsilon^{ab3}\hat{x}^b
\label{ztnhat}
\end{eqnarray}
On the other hand, for the finite T group element in (\ref{pa}), the
relation (\ref{nhat}) leads to
\begin{eqnarray}
\hat{N}^a=\cos\left(\frac{4m\pi t}{\beta}\right)\delta^{a3} +
\left[1-\cos\left(\frac{4m\pi t}{\beta}\right)\right]
\hat{n}^a \hat{n}^3 -
\sin\left(\frac{4m\pi t}{\beta}\right)
\epsilon^{ab3}\hat{n}^b
\label{ftnhat}
\end{eqnarray}
If we consider the particular $CP^1$ instanton $\hat{n}$ to
be the embedding unit vector $\hat{x}$ of ${\bf S}^2$ into ${\bf R}^3$, then we
can compare the two maps $\hat{N}$ in (\ref{ztnhat}) and (\ref{ftnhat}).
The difference, once again, is that the trigonometric factors wrap twice
as often in the finite T case. Thus, geometrically speaking, the
corresponding pre-images link through one another twice as often, thereby
explaining the fact that the winding number of $g$ in (\ref{even}) is an
even integer.

\section{Coset construction}

In this section, we consider an even simpler construction at finite
temperature. Recall, from the studies of thermal Abelian
transformations, that they have the general form
\begin{equation}
U(\bar{x},t) = e^{i(2\pi m {t\over \beta} + \Omega (t,\bar{x}))} =
e^{2i\pi m {t\over\beta}}\,e^{i\Omega (t,\bar{x})}
\end{equation}
where $m$ is an integer and $\Omega$ is periodic and we choose:
\begin{equation}
\Omega(\bar{x},t=0) = \Omega(\bar{x},t=\beta) = 0
\end{equation}
With such a construction, the Abelian group element $U(\bar{x},t)$ is
manifestly periodic, 
and brings out, in a natural and intuitive manner, the windings of the
gauge transformation around the circle along the time direction.

We can try to generalize this to $SU(2)$ in the following
manner. First, let us recall that $SU(2)$ has an Abelian
subgroup. Therefore, let us write an element of $SU(2)$ as
\begin{equation}
g(\bar{x},t) = h(t) u(\bar{x}) = e^{2m\pi i {t\over \beta}\sigma_{3}}
u(\bar{x})
\end{equation}
This is clearly periodic. Now choose the space part
$u(\bar{x})$ of the group element to be
\begin{equation}
u(\bar{x}) = \exp\left[i\pi\,\hat{\bar{x}}\cdot
\bar{\sigma}\,f(\rho)\right]
\end{equation}
where $\hat{\bar{x}}$ is the two-dimensional spatial unit vector,
$\bar{\sigma}$ represents the two spatial Pauli matrices, and
$\rho$ is the two dimensional radial coordinate. Then, it is easily shown
that the winding number, in this case, becomes
\begin{eqnarray}
W[g] & = & {1\over 24\pi^{2}} \int_{0}^{\beta} dt\int
d^{2}x\,\epsilon_{\mu\nu\lambda} {\rm Tr}\left( g^{-1}\partial_{\mu}g
g^{-1}\partial_{\nu}g g^{-1}\partial_{\lambda}g\right)\nonumber\\
 & = & m\left[\cos (2\pi f(0)) - \cos (2\pi f(\infty))\right]
\end{eqnarray}
Thus, if we choose
\begin{equation}
f(\rho) = {n\rho\over 2\sqrt{\rho^{2}+\lambda^{2}}}
\end{equation}
where $\lambda$ is an arbitrary length scale, then,
\begin{equation}
W[g] = m \left(1 - (-1)^{n}\right)
\end{equation}
This shows that the winding number can be nonzero, if $n$ is odd. But
then, once again, the winding number $W[g]$ is necessarily an even
integer. The problem, however, lies in the fact that the group
element has the explicit form
\begin{equation}
g(\bar{x},t) = e^{2m\pi i {t\over \beta} \sigma_{3}} \left(\cos \pi
f(\rho) + i \hat{\bar{x}}\cdot \bar{\sigma}\,\sin \pi f(\rho)\right)
\end{equation}
This shows that, with the above choice of $f(\rho)$, the group element is
independent of angles at spatial infinity, $\rho\rightarrow \infty$, for
any fixed $t$, only if $n$ is an even integer. Therefore, while for even
$n$ we can construct an acceptable group element with the required
isotropy, the corresponding winding number is trivial. On the other
hand, for odd $n$, the group element does give a nontrivial winding
number (albeit even), but the group element does not possess the required
isotropy properties. It is worth noting here that, for odd $n$, even
though the group element is not isotropic at spatial infinity,
$g^{-1}\partial_{\xi}g$ is, where $\xi$ is the polar angle. An
alternate, simple way to see the vanishing of the winding number is to
use formula (\ref{comp}).

\section{Enlarging the group manifold}

The failure of the simplistic construction of the previous section is due
to a mismatch between the base and the target manifolds at finite temperature.
To see this more explicitly, note that an element of
$SU(2)$ can be parameterized as
\begin{equation}
g(\bar{x},t) = \exp\left(i\theta\, \hat{N}\cdot\vec{\sigma}\right)
\label{enlarged}
\end{equation}
where $\theta=\theta (\bar{x},t)$, and $\hat{N} =
\hat{N}(\bar{x},t)$ is a three-dimensional unit vector. (For
example, the ansatz of Eq. (\ref{pa}) is a special case where the $t$ and
$\bar{x}$ dependence is separated: $\theta(\bar{x},t)=\theta (t)$, and
$\hat{N}(\bar{x},t)=\hat{n}(\bar{x})$.) Furthermore, $0\leq \theta
\leq\pi$ and the three components of the unit vector $\hat{N}$ can be
traded for two angular variables, say $\psi,\phi$ with $0\leq \psi \leq
\pi$ and $0\leq\phi\leq 2\pi$. Thus, an arbitrary element of $SU(2)$ can
be parameterized in terms of three angles
\begin{equation}
g (\bar{x},t) = g (\theta,\psi,\phi)\quad,\qquad 0\leq \theta ,\psi
\leq
\pi;\,0\leq \phi \leq 2\pi
\end{equation}
where the coordinate dependence is contained in the angular
parameters. This brings out explicitly the identification of $SU(2)$
with ${\bf S}^{3}$ (of course, we do not need to worry about the center of
the group).

In contrast, the space-time manifold, at finite temperature, is
labeled by $(\bar{x},t)$, where
$0\leq {2\pi t\over \beta} \leq 2\pi$  defines the ${\bf S}^{1}$. For
the  spatial
coordinates we can use polar coordinates $(\rho,\xi)$ where $0\leq \xi
\leq 2\pi$. The radial coordinate $\rho$ can further be identified with an
angular variable through a stereographic projection as
\begin{equation}
{\rho\over \lambda} = 2 \tan {\zeta\over 2}
\end{equation}
so that we can describe the two dimensional space by $(\zeta,\xi)$
with $0\leq \zeta \leq \pi$ and $0\leq \xi \leq 2\pi$ ($(\zeta,\xi)$,
for example, can be thought of as the polar angles of
(\ref{cp1}).). This is  the
${\bf S}^{2}$ associated with the spatial manifold. Thus, we can also
represent the group element as
\begin{equation}
g = g (t,\zeta,\xi)
\end{equation}
and the requirement of asymptotic isotropy can now be written as
\begin{equation}
g(t,\zeta=\pi,\xi) = g_{0}(t)
\end{equation}

The difficulty for the maps from the space-time manifold to the
group manifold is now clear. Namely, the parameters of the space-time
manifold, $(\tau,\zeta,\xi)$, have ranges that are different from those
in the group manifold, $(\theta,\psi,\phi)$; specifically, the coordinate
parameter $\tau$ ranges from $0$ to $2\pi$, while the group parameter
$\theta$ ranges from $0$ to $\pi$. Therefore, for a map with nontrivial
winding to exist, we must somehow enlarge the group space. For example,
if we enlarge the parameter space of the group by defining
\begin{equation}
\tilde{g} (\theta,\psi,\phi) = \left\{\matrix{
g(\theta,\psi,\phi)\quad  {\rm for}\quad  0\leq \theta \leq \pi\cr\cr
g(2\pi -\theta,\pi-\psi,\pi+\phi)\quad  {\rm for}\quad  \pi\leq \theta \leq
2\pi}\right.
\label{doubled}
\end{equation}
then, the ranges of the parameters in the group space as well as the
space-time manifold will match and a meaningful map may exist.

In fact, with this enlargement, the winding number can be easily
calculated to be
\begin{eqnarray}
W[\tilde{g}]& = & {1\over 24\pi^{2}} \int_{0}^{\beta} dt\int d^{2}x\,
\epsilon_{\mu\nu\lambda}{\rm Tr}\left(
\tilde{g}^{-1}\partial_{\mu}\tilde{g}
\tilde{g}^{-1}\partial_{\nu}\tilde{g}
\tilde{g}^{-1}\partial_{\lambda}\tilde{g}\right)\nonumber\\
& = & {1\over 2\pi^{2}} \int_{0}^{\beta} dt \int_{0}^{\pi} d\zeta
\int_{0}^{2\pi} d\xi\,\sin^{2}\theta
\sin\psi\,\left|{\partial (\theta,\psi,\phi)\over \partial
(t,\zeta,\xi)}\right|\nonumber\\
 & = & {1\over 2\pi^{2}} \int_{0}^{2N'\pi} d\theta\,\sin^{2}\theta
\int_{0}^{\pi} d\psi\,\sin \psi \int_{0}^{2N\pi} d\phi\nonumber\\
 & = & 2NN'
\label{even2}
\end{eqnarray}
Here, we have used the periodicity relations of the form
\begin{eqnarray}
\theta({2\pi\over \beta},\zeta,\xi) & = & \theta(0,\zeta,\xi) +
2N\pi\nonumber\\
\psi(t,0,\xi) & = & 0,\;\psi(t,\pi,\xi) = \pi\nonumber\\
\xi(t,\zeta,2\pi) & = & \xi(t,\zeta,0) + 2N'\pi
\end{eqnarray}

Therefore, we see that the enlargement of the parameters in the group
manifold does allow us to construct a group element that leads to a
nontrivial winding. In fact, it leads to an even winding number much
like the one in Eq.(\ref{pa}). There are several ways to understand
this result. First of all, from the doubling of the parameters in the
group space we see that every point in $(t,\zeta,\xi)$ gets mapped
to two points in $(\theta,\psi,\phi)$, which is the main reason for an
even number of winding. A more geometrical way of understanding the
doubling is to note that the doubling in the group space leads to two
${\bf S}^{3}$'s touching at two points - in some sense folding back
like a torus. This, therefore, leads to a nontrivial product space of
two ${\bf S}^{3}$'s. As a result, we can map the ${\bf S}^{1}$ and the
${\bf S}^{2}$ of the space-time manifold to each of these ${\bf
S}^{3}$'s and there are two ways of doing this, which leads to an even
winding number (and these windings are nontrivial because of the
nontrivial nature of the product of two ${\bf S}^{3}$'s). This also
suggests that such gauge transformations are not completely
characterized by $W[g]$, rather they are determined by the two integers
$N,N'$.
The construction in Eq. (\ref{pa}) illustrates this
general approach with $\theta = {m\pi t\over
\beta},\psi=\zeta,\phi=n\phi$ so that $N=m$ and $N'=n$, which explains
why ansatz (\ref{pa}) leads to even winding numbers only.

\section{Construction for arbitrary winding}

While the previous sections explain why the ansatzes (\ref{pa}) and
(\ref{enlarged}), (\ref{doubled}) lead to even winding numbers, these
anstazes still do not accommodate any integer winding number, nor do they
have a smooth zero temperature limit. In this section, we present a
construction that solves both these problems. This new ansatz is
extremely simple, and is motivated by insights from the
$0+1$ dimensional models.

In $0+1$ dimension, at zero temperature, an Abelian {\em large} gauge
transformation can be written as
\begin{equation}
U_{(0+1)}(t) = e^{i\Omega(t)}
\end{equation}
where
\begin{equation}
\Omega(t) =2m\,{\rm arctan}(t)= -im\,\log \left({1+it\over 1-it}\right)
\label{arctan}
\end{equation}
The coefficient $m$ is required to be an integer so that the Abelian group
element
\begin{eqnarray}
U_{(0+1)}(t)=\left({1+it\over 1-it}\right)^m
\label{analytic}
\end{eqnarray}
is analytic in $t$. This is why the winding number
\begin{eqnarray}
W[U]=\frac{1}{2\pi}\int dt \,\frac{d\Omega}{dt}=m
\label{01winding}
\end{eqnarray}
is an integer. Now, to extend this to finite temperature, we can
compactify the time direction with the redefinition of variables,
\begin{equation}
t \rightarrow \tan {\pi t\over \beta}
\label{compactify}
\end{equation}
so that the transformed time spans $-{\beta\over 2}\leq t \leq
{\beta\over 2}$. This changes the gauge transformation parameter $\Omega$
in (\ref{arctan}) to
\begin{eqnarray}
\Omega  =  -im\,\log \left({1+i\tan {\pi t\over \beta}\over
1-i\tan {\pi t\over \beta}}\right)={2m\pi t\over \beta}
\end{eqnarray}
Then $U=\exp(2m\pi i t/\beta)$ is precisely the familiar thermal Abelian
{\em large} gauge transformation. In this compactified language, $m$ must
be an integer in order to preserve the correct periodicity properties of
matter fields to which the Abelian gauge field is coupled.

It is straightforward to generalize this idea to a non-Abelian gauge
group in $2+1$ dimensions. For example, for $SU(2)$ we take the zero
temperature ansatz (\ref{zt}) and map $x^3=t\rightarrow \tan(\pi t/\beta)$
as in (\ref{compactify}). This suggests the following ansatz:
\begin{eqnarray}
g(\bar{x}, t)=\exp\left(m\pi i\left[
\frac{\frac{\beta}{\pi}\tan(\frac{\pi t}{\beta})\,\sigma_3
+\bar{x}\cdot\bar{\sigma}}{\sqrt{{\beta^{2}\over \pi^{2}} \tan^{2}
{\pi t\over \beta} +
\bar{x}^2 + \lambda^{2}}}\right]\right)
\label{answer}
\end{eqnarray}
where $m$ is an integer and $\lambda$ is an arbitrary length scale. 

The Euclidean time axis has been compactified to $-{\beta\over 2}\leq
t \leq {\beta\over 2}$, and this group element is clearly
periodic on this interval:
\begin{eqnarray}
g(\bar{x}, t=-\frac{\beta}{2})=g(\bar{x}, t= \frac{\beta}{2})
\label{check}
\end{eqnarray}
Furthermore, in the zero temperature limit, $\beta\to\infty$, the time
coordinate covers the full range from $-\infty$ to $+\infty$, and the
group element (\ref{answer}) reduces smoothly to the zero temperature
{\em large} gauge transformation (\ref{zt}). It is  a straightforward exercise
to check that the winding number of the gauge transformation
(\ref{answer}) is
\begin{eqnarray}
W[g] & = & {1\over 24\pi^{2}} \int_{-{\beta\over 2}}^{\beta\over 2}
dt \int d^{2}x\,\epsilon_{\mu\nu\lambda}  
{\rm Tr}\left(g^{-1}\partial_{\mu}g g^{-1}\partial_{\nu}g
g^{-1}\partial_{\lambda}g\right)\nonumber\\
  & = & m
\end{eqnarray}
which can be {\it any} arbitrary integer. So the winding number is
nontrivial, and is not restricted to just even integers. This result can
be seen in two different ways. First, it is worth pointing out that the
nontrivial contribution, in an explicit calculation of the integrals,
comes from the radial surface rather than the temporal direction.
Alternatively, a very simple way to
see that the winding number is $m$ is to note that the finite
temperature ansatz is obtained from the zero temperature ansatz through a
coordinate redefinition and, since the winding number is the integral of a
three form, $W[g]=\frac{1}{24\pi^2}\int Tr((g^{-1}dg)^3)$, it is invariant
under this coordinate redefinition. We, therefore, conclude that the simple
ansatz (\ref{answer}) leads to a general non-Abelian thermal {\em large}
gauge transformation.

Our ansatz in (\ref{answer}) is also partly motivated by the well-known
form of the Harrington-Shepard caloron solutions \cite{caloron,gpy} which
are instantons in 4 dimensional Yang-Mills theory that are periodic in
Euclidean time. These also have a trigonometrically compactified time
coordinate, but the calorons are much more complicated than the finite
temperature large gauge transformations constructed in (\ref{answer})
because the calorons are required to satisfy the non-trivial self-dual
Yang-Mills equations after compactification. The large gauge
transformation group element $g$ need not satisfy any particular
differential equations; it simply is required to have a nontrivial
winding number.

It is worth comparing the nature of the explicit ansatzes in
Eqs. (\ref{pa}) and (\ref{answer}), at least for even winding
numbers. Let us note the following properties of the ansatz in
Eq. (\ref{pa}). For any fixed $t$, this group element defines a two
sphere, ${\bf S}^{2}$, while for any fixed coordinate, it defines a
circle, ${\bf S}^{1}$. On the other hand, for even $m$, say $m=2$, we
can write the ansatz of Eq. (\ref{answer}) also in a similar form, namely,
\begin{equation}
g(\bar{x},t) = \exp\left(2\pi i \theta\,
\hat{\theta}\cdot\vec{\sigma}\right)
\end{equation}
where $\theta$ is the magnitude and $\hat{\theta}$ the unit vector for
the three-vector:
\begin{eqnarray}
\vec{\theta}&=&\left(\bar{x},\frac{\beta}{\pi}\tan\frac{\pi t}{\beta}
\right)\nonumber\\
\theta &=& \left(1 - {\lambda^{2}\over {\beta^{2}\over \pi^{2}} \tan^{2}
{\pi t\over \beta} + \bar{x}^{2} + \lambda^{2}}\right)^{1\over 2}
\end{eqnarray}
Clearly, $0\leq \theta \leq 1$ and this looks qualitatively similar to
the ansatz in Eq. (\ref{pa}). However, in the present
case, $\hat{\theta} = \theta_{a}(\bar{x}, t)$ depends on both the
spatial and temporal coordinates, while the unit vectors
$\hat{n}(\bar{x})$, in Eq. (\ref{pa}) depend only on the spatial
coordinates $\bar{x}$. Also, although for any fixed $t$, the present
ansatz describes a two sphere, it does not describe a circle for any
fixed spatial coordinates. In fact, note that $\theta$ attains its
maximum value $1$ only at the points $t =\pm {\beta\over 2}$ or
$r=\infty$. Finally, we note that the construction in (39) is not
completely analytic in the sense that the second derivative, with
respect to $t$, of the group
element has a discontinuity at $t=\beta/2$. This non-analyticity may be
the price one has to pay in order to get odd values for the winding
number.

\section{Conclusion}

We conclude briefly by commenting that we have presented a very simple
ansatz (\ref{answer}) for finite temperature {\em large} gauge transformations
in $SU(2)$. The generalization to other compact gauge groups is
straightforward, just like at zero temperature. Our ansatz has a smooth
zero temperature limit, and can accommodate any integer winding number.
We have also discussed the properties of other ansatzes for finite
temperature {\em large} gauge transformations, and given several complementary
geometric interpretations of why the ansatz (\ref{pa}) only produces even
winding number. Given our ansatz, it will be interesting to study how the
$2+1$ dimensional parity-odd effective action responds to these genuinely
non-Abelian finite temperature {\em large} gauge transformations.

\bigskip
\noindent{\bf Acknowledgment:}  AD would like to thank Prof. S. Okubo for
many helpful discussions, and GD thanks R. MacKenzie for helpful
correspondence. This work was supported in part by US DOE grant numbers
DE-FG02-92ER40716, DE-FG-02-91ER40685 and by CNPq (Brasil). 

\appendix

\section{Some properties of {\em large} gauge transformations:}

In this appendix, we collect some formulae involving the winding
numbers that simplify some of the calculations.
Given a gauge transformation, $g$, the winding number is defined as
\begin{equation}
W[g] = {1\over 24\pi^{2}} \int d^{3}x\,{\rm Tr}\,\epsilon_{\mu\nu\lambda}
g^{-1}\partial_{\mu}g g^{-1}\partial_{\nu}g g^{-1}\partial_{\lambda}g
\end{equation}
A {\em small} gauge transformation is one for which the winding number
vanishes (namely, it is contractible to identity), while the ones with
a nontrivial winding number are known as {\em large} gauge
transformations. 

Winding numbers are additive, namely, if $f = gh$,
then (with appropriate asymptotic fall off),
\begin{equation}
W[f] = W[gh] = W[g] + W[h]
\end{equation}
This can be seen in a simple manner as follows. Let us define
\begin{equation}
X_{\mu} = g^{-1}\partial_{\mu}g,\qquad Y_{\mu} =
\partial_{\mu}h h^{-1},\qquad Z_{\mu} = f^{-1}\partial_{\mu}f =
h^{-1}(X_{\mu}+Y_{\mu})h 
\end{equation}
It can now be easily seen that
\begin{equation}
\epsilon_{\mu\nu\lambda} {\rm Tr}\,(Z_{\mu}Z_{\nu}Z_{\lambda}) =
\epsilon_{\mu\nu\lambda} {\rm Tr}\left[X_{\mu}X_{\nu}X_{\lambda} +
Y_{\mu}Y_{\nu}Y_{\lambda} - 3
\partial_{\lambda}(X_{\mu}Y_{\nu})\right]
\end{equation}
It follows now that, if $X_{\mu},Y_{\mu}$ vanish sufficiently rapidly
for asymptotic distances, the third term would vanish upon integration
and we have
\begin{equation}
W[gh] = W[g] + W[h]\label{comp}
\end{equation}
This also implies that
\begin{equation}
W[g^{-1}] = - W[g]
\end{equation}

Let us next define gauge transformations belonging to a class as a set
of gauge transformations, which can be continuously deformed to one
another. Thus, for example, two transformations, $g_{1},g_{2}$ belong
to the same class, if there exists a set of gauge transformations
(suppressing the dependence on coordinates)
\begin{equation}
g(a;\bar{x},t) \equiv g(a)
\end{equation}
depending smoothly on a parameter \lq $a$' (which may in reality stand
for a set of parameters), such that
\begin{equation}
g(a_{1}) = g_{1},\qquad g(a_{2}) = g_{2}
\end{equation}
The class of {\em small} gauge transformations, then, corresponds to a
special set of transformations, with the property that there exists a
parameter $a_{0}$ for which
\begin{equation}
g(a_{0}) = {\bf 1}
\end{equation}

Given a set of gauge transformations, $g(a)$, let us construct from
these a one parameter family of gauge transformations as
\begin{equation}
f(a_{1},a) = g^{-1}(a_{1}) g(a)
\end{equation}
where we assume that $a_{1}$ has a fixed value and $a$ is
variable. Then, clearly,
\begin{equation}
f(a_{1},a_{1}) = {\bf 1}
\end{equation}
Consequently, it follows that $f(a_{1},a)$ defines a class of {\em
small} gauge transformations (for which the winding number
vanishes). It follows now that
\begin{equation}
W[g(a)] = W[g(a_{1})]
\end{equation}
Namely, every member of the set $g(a)$ belonging to a class of {\em
large} gauge transformations have the same winding number and they
differ from one another only by {\em small} gauge transformations.

Let us define (in connection with a $2+1$ dimensional thermal theory)
\begin{equation}
\rho^{2} = x^{2} + y^{2},\qquad r^{2} = \rho^{2} + t^{2}
\end{equation}
Then, as we have seen, for a fixed time, asymptotic isotropy implies
\begin{equation}
g(\bar{x},t) \rightarrow g_{0}(t)\quad {\rm as}\quad \rho\rightarrow
\infty
\end{equation}
Since $g_{0}(t)$ is periodic, it describes a map from a circle to
$SU(2)$ (${\bf S}^{3}$). Such a map is trivial since a circle on ${\bf
S}^{3}$ is contractible to a point. Therefore, $g_{0}(t)$ represents a
{\em small} gauge transformation (An alternate way of seeing this is to note
that it depends only on the time coordinate and, consequently, the
winding number must vanish.). It follows that we can always define a new
transformation 
\begin{equation}
\tilde{g} (\bar{x},t) = g_{0}^{-1}(t) g(\bar{x},t)
\end{equation}
which belongs to the same class and has the simpler asymptotic form
\begin{equation}
\tilde{g} (\bar{x},t) \rightarrow {\bf 1},\quad {\rm as} \quad
\rho\rightarrow \infty
\end{equation}

Let us next look at this transformation at a fixed time, say $t_{0}$,
namely, $\tilde{g}(\bar{x},t_{0})$. This defines a map from ${\bf S}^{2}$ to
the group $SU(2)$ (${\bf S}^{3}$). This is also a trivial map (since
it does not depend on the time coordinate, the winding number is
zero). Therefore, we can define a gauge transformation
\begin{equation}
\bar{g}(\bar{x},t) = \tilde{g}^{-1} (\bar{x},t_{0}) \tilde{g}(\bar{x},t)
\end{equation}
which will be in the same class as $g,\tilde{g}$. Furthermore, it will have
the property that
\begin{equation}
\bar{g}(\bar{x},t_{0}) = {\bf 1},\qquad \bar{g}(\bar{x},t) \rightarrow {\bf
1}\quad {\rm as}\quad \rho\rightarrow \infty
\end{equation}
This can be thought of as the generalization of the boundary condition
at zero temperature to finite temperature. Namely, at zero
temperature, the boundary condition corresponds to choosing
$\bar{g}\rightarrow {\bf 1}$ as $r\rightarrow \infty$. At finite
temperature, on the other hand, we can think of the boundary of
space-time to be at $t=t_{0}$ and $\rho\rightarrow \infty$.


\begin{thebibliography}{99}

\bibitem{dunne} G. V. Dunne, K. Lee and C. Lu, ``Finite temperature
Chern-Simons coefficient'', Phys. Rev. Lett. {\bf 78} (1997) 3434,
(hep-th/9612194).

\bibitem{deser1} S. Deser, L. Griguolo and D. Seminara, ``Gauge
Invariance, Finite Temperature and Parity Anomaly in D=3'', Phys. Rev.
Lett. {\bf 79} (1997) 1976, (hep-th/9705052); ``Effective QED Actions:
Representations, Gauge Invariance, Anomalies and Mass Expansions'', Phys.
Rev. D {\bf 57} (1998) 7444, (hep-th/9712066); ``Definition of
Chern-Simons Terms in Thermal QED in Three-dimensions Revisited'',
Commun. Math. Phys. {\bf 197} (1998) 443, (hep-th/9712132).

\bibitem{schaposnik}  C. D. Fosco, G. Rossini and F. A. Schaposnik,
``Induced Parity Breaking Term at Finite Temperature'', Phys. Rev. Lett.
{\bf 79} (1997) 1980, (Erratum), ibid {\bf 79} (1997) 4296,
(hep-th/9705124); ``Abelian and Non-Abelian Induced Parity Breaking Terms
at Finite Temperature'', Phys. Rev. D {\bf 56} (1997) 6547,
(hep-th/9707199).

\bibitem{lhl} For a review, see G. V. Dunne, ``Aspects of Chern-Simons
Theory'', 1998 Les Houches Lectures, in the Proceedings, {\it Topological
Aspects of  Low Dimensional Systems}, A. Comtet et al (Editors), 
(Springer-Verlag, 2000), (hep-th/9902115). 

\bibitem{deser2} S. Deser, R. Jackiw and S. Templeton, ``Topologically
massive gauge theories'', Ann. Phys. {\bf 140} (1982) 372.

\bibitem{redlich} A. N. Redlich, ``Gauge Noninvariance and Parity
Violation of Three-Dimensional Fermions'', Phys. Rev. Lett. {\bf 52}
(1984) 18, ``Parity Violation and Gauge Noninvariance of the Effective
Gauge Field Action in Three Dimensions'', Phys. Rev. D {\bf 29} (1984)
2366.

\bibitem{pr} R. D. Pisarski and S. Rao, ``Topologically massive
chromodynamics in the perturbative regime'', Phys. Rev. D {\bf 32} (1985)
2081.

\bibitem{pisarski} R. D. Pisarski, ``Topologically massive chromodynamics
at finite temperature'', Phys. Rev. D {\bf 35} (1987) 664.


\bibitem{bdf} F. Brandt, A. Das and J. Frenkel, ``Induced parity violating
thermal effective action for (2+1)-dimensional fermions interacting with a
non-Abelian background'', (hep-th/0107120); F. Brandt, A. Das, J. Frenkel
and J. C. Taylor, ``Derivative expansion and the parity violating
effective action for thermal (2+1)-dimensional QED at higher orders'',
Phys. Rev. D {\bf 64} (2001) 065018, (hep-th/0103221). 

\bibitem{roman1} R. Jackiw, ``Introduction to the Yang-Mills
Quantum Theory'', Rev. Mod. Phys. {\bf 52} (1980) 661.

\bibitem{rajaraman} R. Rajaraman, {\it Solitons and Instantons},
(North-Holland, New York, 1982).

\bibitem{comment} Actually, in the presence of just adjoint fields, $g$ 
can be periodic up to multiplication by an element of the center of the
gauge group, which for $SU(2)$ would mean $g$ being either periodic or
anti-periodic. This has interesting implications for monopoles in $2+1$
dimensions: R. D. Pisarski, ``Magnetic monopoles in topologically
massive gauge theory'', Phys. Rev. D {\bf 34} (1986) 385. But
with general fields such as fundamental matter fields, $g$ must be
strictly periodic.

\bibitem{kapusta} J. Kapusta, {\it Finite Temperature Field Theory}, 
(Cambridge University Press, 1989).

\bibitem{lebellac} M. Le Bellac, {\it Thermal Field Theory}, (Cambridge
University Press, 1996). 

\bibitem{dasbook} A. Das, {\it Finite Temperature Field Theory} (World
Scientific, 1997).

\bibitem{niemi} E. Langmann and A. Niemi, ``Towards a string
representation of infrared SU(2) Yang-Mills Theory'', (hep-th/9905147).

\bibitem{paul} R. A. Battye and P. M. Sutcliffe, ``Solitons, links and
knots'', Proc. R. Soc. Lond. {\bf A 455} (1999) 4305, (hep-th/9811077).

\bibitem{tsutsui} T. Tsurumaru, I. Tsutsui and A. Fujii, ``Instantons, 
Monopoles and the Flux Quantization in the Faddeev-Niemi
Decomposition'', Nucl.Phys. B {\bf 589} (2000) 659, (hep-th/0005064).

\bibitem{roman} R. Jackiw and S-Y. Pi, ``Creation and evolution of
magnetic helicity'', Phys. Rev. D {\bf 61} (2000) 105015,
(hep-th/9911072).

\bibitem{adam} C. Adam, B. Muratori and C. Nash, ``Multiple
zero modes of the Dirac operator in three-dimensions'', Phys. Rev. D {\bf
62} (2000) 085026, (hep-th/0001164); ``Particle creation via
relaxing hypermagnetic knots'', Phys. Rev. D {\bf 62} (2000) 105027,
(hep-th/0006230 ); ``Hopf instantons and the Liouville equation in target
space'', Phys. Lett. B {\bf 479} (2000) 329, (hep-th/0001163). 

\bibitem{jahn} O. Jahn, ``Instantons and monopoles in general Abelian
gauges'', J. Phys. A: Math. Gen. {\bf 33} (2000) 2997, (hep-th/9909004).

\bibitem{vanbaal} P. van Baal and A. Wipf, ``Classical gauge vacua as
knots'', Phys. Lett. B {\bf 515} (2001) 181, (hep-th/0105141). 

\bibitem{caloron} B. J. Harrington and H. K. Shepard, ``Periodic
Euclidean solutions and the finite-temperature Yang-Mills gas'', Phys.
Rev. D {\bf 17} (1978) 2122.

\bibitem{gpy} D. J. Gross, R. D. Pisarski and L. G. Yaffe, ``QCD and
instantons at finite temperature'', Rev. Mod. Phys. {\bf 53} (1981) 43.




\end{thebibliography}
\end{document}